\documentclass[runningheads]{llncs}
\usepackage[T1]{fontenc}
\usepackage{graphicx}
\usepackage{amsmath}
\usepackage{amsfonts}
\usepackage{amssymb}
\usepackage{booktabs}
\usepackage{float}
\usepackage{graphicx}
\usepackage{hyperref}
\usepackage{cleveref}
\usepackage{adjustbox}
\usepackage{subcaption}
\usepackage{caption}
\usepackage{geometry}
\usepackage{multirow}
\usepackage{enumerate}
\usepackage{array}

\begin{document}
\title{ggskewboxplots: Enhanced Boxplots for Skewed Data in R}
\titlerunning{ggskewboxplots}
%
\author{Mustafa Cavus\orcidID{0000-0002-6172-5449}}
%
%
\institute{Department of Statistics, Eskisehir Technical University, Turkiye\\
\email{mustafacavus@eskisehir.edu.tr}}
%
\maketitle             
\begin{abstract}
Traditional boxplots are widely used for summarizing and visualizing the distribution of numerical data, yet they exhibit significant limitations when applied to skewed or heavy-tailed distributions, often leading to misclassification of outliers through \textit{swamping}---flagging typical observations as outliers or \textit{masking}---failing to detect true outliers. This paper addresses these limitations by systematically evaluating several alternative boxplot methods specifically designed to accommodate distributional asymmetry. We introduce \texttt{ggskewboxplots}, an R package that integrates multiple robust and skewness-aware boxplot variants, providing a unified and user-friendly framework for exploratory data analysis. Using extensive Monte Carlo simulations under controlled skewness and kurtosis conditions, implemented via the mosaic approach based on the Skewed Exponential Power distribution, we assess the sensitivity and specificity of each method. Simulation results indicate that classical Tukey-style boxplots are highly prone to swamping and masking, whereas robust skewness-adjusted variants—particularly those leveraging quartile-based skewness measures or medcouple-based adjustments—achieve substantially better performance. These findings offer practical guidance for selecting reliable boxplot methods in applied settings and demonstrate how the \texttt{ggskewboxplots} package facilitates accessible, distribution-aware visualizations within the familiar \texttt{ggplot2} workflow.

\end{abstract}
\newpage
\section{Introduction}

Boxplots are widely used statistical tools for visualizing the distribution of a variable. First introduced by Spear \cite{spear1952} and later popularized by Tukey \cite{tukey1977exploratory}, they provide essential summary statistics such as the median, quartiles, and whiskers, while highlighting potential outliers in a dataset. Due to their simplicity, interpretability, and usefulness in exploratory data analysis, boxplots remain a common tool among researchers and practitioners \cite{benjamini1988boxplot,sim2005outlier}. In practice, these plots can be readily produced in both base R \cite{baseR}, e.g., \texttt{boxplot()}, and the \texttt{ggplot2} package \cite{ggplot2}, e.g., \texttt{geom\_boxplot()}, which together constitute the primary tools for routine boxplot visualization.

However, a fundamental limitation of the traditional (aka Tukey-style) boxplot is its reliance on assumptions of symmetry and light-tailedness. It defines outliers as observations falling beyond $1.5$ times the interquartile range from the lower and upper quartiles. While this rule performs well under symmetric, mesokurtic distributions like the normal, it tends to over-flag regular observations as outliers — or fail to detect genuine outliers — when the data are skewed or heavy-tailed \cite{hubert2008adjusted,bruffaerts2014generalized}. These issues correspond to common problems in outlier detection known as \textit{swamping} and \textit{masking}. \textit{Swamping} occurs when typical (non-outlying) observations are incorrectly identified as outliers, leading to a decrease in the specificity (true negative rate). Conversely, \textit{masking} happens when actual outliers remain undetected because they are obscured by the overall data distribution, reducing sensitivity (true positive rate). For example, in a right-skewed dataset, the classic boxplot may flag many normal observations on the long tail as outliers (\textit{swamping}), while simultaneously missing subtle but important extreme values (\textit{masking}). These limitations highlight the need for distribution-aware alternatives that maintain both high sensitivity and specificity across diverse data shapes.

To address these shortcomings, several alternative boxplots have been proposed. One notable example is the adjusted boxplot \cite{hubert2008adjusted}, which utilizes the robust medcouple to adjust whisker lengths according to the skewness. Other approaches involve log-transformations \cite{schwertman2004boxplot}, extreme value theory \cite{bhattacharya2023outlier}, or fully nonparametric variants \cite{dovoedo2015boxplot,walker2018improved}.

Beyond methodological developments, the growing flexibility of statistical software has also broadened the practical landscape for boxplot customization. In \texttt{R}, both base graphics and the \texttt{ggplot2} ecosystem allow users to override or fully redefine the statistical components of a boxplot—such as whisker limits, quartile definitions, or outlier rules—through user-supplied summaries or custom statistical layers. For example, the functions \texttt{bxp()} and \texttt{ggproto-}based extensions make it possible to construct boxplots using arbitrary formulas rather than the classical Tukey definition. However, despite this technical flexibility, the statistical logic of a boxplot cannot be modified directly by end users, and manually supplying the required summaries for each alternative method can be cumbersome—especially for users unfamiliar with the diverse boxplot variants proposed in the literature. As a result, many robust or distribution-aware boxplot methods remain underutilized in applied work, highlighting the need for accessible tools that implement these approaches in a unified and user-friendly manner.

This paper addresses the lack of an integrated R package that provides multiple alternative boxplot methods designed to handle skewed and heavy-tailed data, aiming to offer a practical and user-friendly tool for data analysts facing distributional challenges. To this end, we developed a new R package implementing a range of robust and flexible boxplot methods. Additionally, we systematically compare the performance of these alternative boxplots across varying levels of skewness and kurtosis, guiding users in selecting the most appropriate method regardless of the underlying data distribution. Extensive Monte Carlo simulations are conducted under controlled distributional settings to assess the sensitivity (true positive rate) and specificity (true negative rate) of each method. For performance evaluation, we employ the mosaic approach \cite{sanchez2018visualizing}, based on the Skewed Exponential Power Distribution \cite{zhu2009properties}, which enables flexible control over both skewness and tail weight. We chose this approach because, unlike costly simulation designs that rely on multiple separate distributions, it can represent a wide range of distribution families within a unified framework, thereby enhancing the efficiency and generalizability of our analysis. The main contributions of this paper can be summarized as follows:
\begin{enumerate}[(i)]
    \item We introduce \texttt{ggskewboxplots}, an \texttt{R} package that integrates multiple alternative boxplot methods tailored for skewed and heavy-tailed data, providing a unified and user-friendly framework for practitioners.
    \item We conduct extensive Monte Carlo simulations under controlled skewness and kurtosis settings to evaluate the sensitivity and specificity of each boxplot method, offering practical guidance on method selection depending on the underlying data distribution.
    \item We demonstrate the advantages of these alternative boxplots in mitigating swamping and masking effects, highlighting their practical relevance for exploratory data analysis and outlier detection in diverse datasets.
\end{enumerate}

The remainder of this paper is structured as follows. Section 2 introduces the alternative boxplot methods included in the study. Section 3 describes the simulation design and evaluation metrics. Section 4 presents the results, followed by discussion and conclusions in Section 5.
\section{Methodology}          
\begin{figure}[h!]
    \centering
    \includegraphics[width=0.75\linewidth]{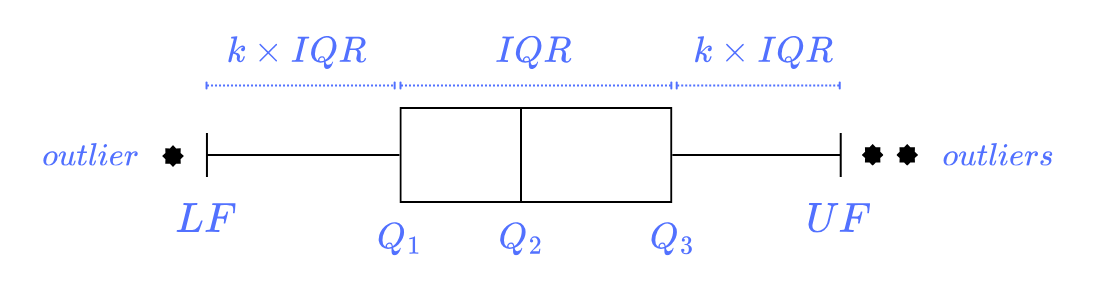}
    \caption{Illustration of a classical Tukey-style boxplot. Observations outside the lower and upper fences $[LF, UF] = [Q_1 - k \cdot \mathrm{IQR}, \, Q_3 + k \cdot \mathrm{IQR}]$ are marked as outliers.}
    \label{fig:boxplot}
\end{figure}

The classical Tukey boxplot defines outliers as those observations falling below  
$Q_1 - k \cdot \mathrm{IQR}$ or above $Q_3 + k \cdot \mathrm{IQR}$, where $Q_1$, $Q_2$, and $Q_3$ denote the first, second (median), and third quartiles, respectively, and $\mathrm{IQR} = Q_3 - Q_1$ is the interquartile range, as illustrated in Figure~\ref{fig:boxplot}. Here, $k$ is a coefficient and is called the whisker coefficient. It controls the size of the fences and is taken as $1.5$ mostly in practice. The values are higher or lower than these fences, labeled as outliers. This rule is widely used in practice to identify the outliers in the data due to its simplicity.

However, several modifications have been proposed to accommodate skewed or heavy-tailed data better. These adaptations aim to improve outlier detection performance by adjusting the whisker length or thresholding rules based on distributional characteristics such as skewness and kurtosis. In the following section, we summarize a set of distribution-aware boxplot variations that address these issues.

\subsection{Kimber's Boxplot}

Kimber \cite{kimber1990exploratory} introduced a variation of Tukey’s fences that accommodates asymmetry between the lower and upper tails. Instead of relying on a single interquartile range, the approach divides it at the median, resulting in two distinct halves: the lower semi-interquartile range ($SIQR_L = Q_2 - Q_1$) and the upper semi-interquartile range ($SIQR_U = Q_3 - Q_2$). These tailored ranges enable the construction of fences that incorporate the skewness present within the interquartile range.

\begin{equation}
    Q_1 - 2k \cdot (Q_2 - Q_1), \quad Q_3 + 2k \cdot (Q_3 - Q_2).
\end{equation}

\noindent This approach maintains the simplicity of the original boxplot while allowing the whisker lengths to reflect asymmetry in the distances between quartiles.

\subsection{Hubert's Boxplot}

Hubert and Vandervieren \cite{hubert2008adjusted} proposed an adjusted boxplot that modifies the whiskers using the \emph{medcouple} ($MC$), a robust measure of skewness:

\begin{equation}
    Q_1 - k \cdot \exp(-3MC) \cdot \mathrm{IQR}, \quad Q_3 + k \cdot \exp(3MC) \cdot \mathrm{IQR}
\end{equation}

\noindent where $MC$ is defined as a robust median difference of right- and left-centered data: 

\begin{equation}
    MC = \frac{\mathrm{median}(x_i > Q_2) - \mathrm{median}(x_i < Q_2)}{\mathrm{IQR}}. 
\end{equation}

\noindent This adjustment results in tighter or wider whiskers depending on the direction and magnitude of the skewness, providing better control over false outlier labeling.

\subsection{Adil’s Boxplot}

Adil et al. \cite{adil2015modified} proposed a refinement of the adjusted boxplot by incorporating both the medcouple ($MC$) and the classical moment-based skewness ($SK$) into the fence construction. While the earlier method of Hubert's Boxplot \cite{hubert2008adjusted} applied fixed constants as exponents scaled by the medcouple to control the asymmetry of the whiskers, Adil's Boxplot replaces these constants with the product of the absolute medcouple and the sample skewness.

The motivation behind this adjustment lies in the behavior of the classical skewness measure: when the skewness is small, the influence on the fence is minimal, resulting in tighter bounds; conversely, larger skewness values expand the fence width, allowing the method to respond to varying degrees of asymmetry in the data adaptively. The modified fences are expressed as:

\begin{equation}
    Q_1 - k \cdot \mathrm{IQR} \cdot \exp(SK \cdot |MC|), \quad Q_3 + k \cdot \mathrm{IQR} \cdot \exp(SK \cdot |MC|)
\end{equation}

\noindent where $SK$ is the adjusted skewness statistic, bounded to lie within $[-3.5, 3.5]$ to prevent excessive stretching of the whiskers for extremely skewed distributions. This capping mechanism ensures that the boxplot remains sensitive enough to detect outliers in moderately skewed data without producing overly wide fences that could mask them.

By combining the robustness of the $MC$ with the interpretability of classical skewness, this method offers a balanced approach to skewness adjustment. In particular, it mitigates the risk of Type II errors in outlier detection, which can occur when overly generous fences fail to flag extreme observations in moderately skewed samples.

\subsection{Babura’s Boxplot}

Babura et al. \cite{babura2017modified} used \emph{Bowley's coefficient of skewness} ($B_C$) to adjust the whisker lengths:

\begin{equation}
    Q_1 - k \cdot \exp(6 \cdot B_C) \cdot \mathrm{IQR}, \quad Q_3 + k \cdot \exp(6 \cdot B_C) \cdot \mathrm{IQR}
\end{equation}

\noindent where $B_C$ is given by 

\begin{equation}
    B_C = \frac{Q_3 + Q_1 - 2Q_2}{\mathrm{IQR}}.
\end{equation}

\noindent Here, $B_C$ captures asymmetry in the quartile positions and is bounded between $-1$ and $1$.

\subsection{Walker’s Boxplot}

Walker et al. \cite{walker2018improved} proposed a more flexible variation of the classical Tukey boxplot by introducing a skewness-sensitive formulation. Building on Kimber's Boxplot \cite{kimber1990exploratory} idea of using semi-interquartile ranges to define the lower and upper fences asymmetrically, Walker and colleagues incorporated Bowley’s coefficient of skewness to enable a more dynamic adjustment.

$B_C$ is the quartile-based measure of asymmetry that quantifies the degree and direction of skewness in a distribution. It is defined as the difference between the upper and lower semi-interquartile ranges, normalized by their sum. It ranges from $-1$ to $1$, where values near zero indicate a symmetric distribution, and positive or negative values reflect right or left skewness, respectively. A greater absolute value of $B_C$ indicates stronger skewness. This coefficient is directly used in the construction of the boxplot fences, using it as a multiplicative adjustment factor. The proposed functional form is as follows:

\begin{equation}
    Q_1 - k \cdot \text{IQR} \cdot \frac{1 - B_C}{1 + B_C}, \quad Q_3 + k \cdot \text{IQR} \cdot \frac{1 + B_C}{1 - B_C}
\end{equation}

This structure allows for more pronounced fence adjustments when $B_C$ deviates substantially from zero. For instance, in the case of right-skewed data where $B_C > 0$, the factor $(1 - B_C)/(1 + B_C)$ becomes less than one. Consequently, the lower fence is extended further outward while the upper fence contracts, leading to a more adaptive outlier detection in skewed distributions.

Walker’s method retains the classical Tukey fences for approximately symmetric data, while offering a more realistic delineation of extreme values when skewness is present. This adaptability is especially valuable in domains such as finance or healthcare, where skewed distributions are common and robust outlier detection is critical.

\subsection{Junsawang’s Boxplot}

Junsawang et al. \cite{junsawang2021robust} proposed a dynamic modification of Hubert's Boxplot \cite{hubert2008adjusted} that incorporates the ratio of semi-interquartile ranges into the exponent:

\begin{equation}
    Q_1 - k \cdot \exp\left(B_C \cdot \frac{Q_2 - Q_1}{Q_3 - Q_2}\right) \cdot \mathrm{IQR}, \quad Q_3 + k \cdot \exp\left(B_C \cdot \frac{Q_2 - Q_1}{Q_3 - Q_2}\right) \cdot \mathrm{IQR}
\end{equation}

\noindent This approach flexibly adjusts the whiskers according to local asymmetry around the median, providing robustness across a wide variety of skewed data scenarios.

\section{Package Structure} 
The \texttt{ggskewboxplots} is designed to extend the functionality of \texttt{ggplot2} by providing skewness-aware boxplot layers. The main function of the package is \texttt{geom\_skewboxplot()}, which acts as a drop-in replacement for \texttt{geom\_boxplot()} but incorporates several alternative methods for adjusting whisker lengths based on data skewness. The statistical transformation logic is handled by a custom \texttt{ggproto} object named \texttt{StatSkewBoxplot}, which computes five-number summaries and outlier boundaries depending on the selected method, e.g., Tukey, Adil, Hubert, etc. The computation relies on the internal helper function \texttt{compute\_skew\_stats()}, which implements multiple formulae for adjusting whiskers depending on robust measures of asymmetry.

In addition to visualization, the package includes a summarization utility, \texttt{summarise\_skewbox()}, which returns a tidy table of modified boxplot statistics. This function is useful for exploratory data analysis or generating custom visual summaries without plotting.

The package is implemented entirely in base R and the tidyverse ecosystem, relying on \texttt{ggplot2} for graphical rendering, and \texttt{dplyr}, \texttt{tidyr}, and \texttt{rlang} for data manipulation and tidy evaluation. Table~\ref{tab:structure} summarizes the core user-facing and internal functions, highlighting their roles within the package.

\begin{table}[h!]
\centering
\caption{An overview of the functions in \texttt{ggskewboxplots}}
\label{tab:structure}
\begin{tabular}{p{4cm}p{9cm}}
\toprule
\textbf{Function} & \textbf{Purpose} \\
\midrule
\texttt{geom\_skewboxplot()} & Main user-facing function; adds a skew-aware boxplot layer to \texttt{ggplot2}. \\
\texttt{compute\_skew\_stats()} & Internal helper; computes five-number summary and whisker bounds. \\
\texttt{summarise\_skewbox()} & Returns modified boxplot statistics in a tidy tibble format. \\
\bottomrule
\end{tabular}
\end{table}

\section{Use Cases} 
The \texttt{ggskewboxplots} package offers a practical and flexible interface for visualizing asymmetric data distributions using skew-aware boxplots. The primary function, \texttt{geom\_skewboxplot()}, seamlessly integrates with the \texttt{ggplot2} ecosystem, allowing users to replace \texttt{geom\_boxplot()} with minimal adjustments, while benefiting from distribution-aware whisker computations.

\subsection{Basic Usage}

The simplest use case involves applying one of the available methods, e.g., \texttt{adil}, \texttt{hubert}, \texttt{walker}, directly to a boxplot:


\begin{figure}[ht]
    \centering
    \begin{minipage}{0.45\textwidth}
    \begin{verbatim}
> library(ggplot2)
> library(ggskewboxplots)
> ggplot(mpg, aes(x = class, y = hwy)) +
+  geom_skewboxplot(method = "walker") +
+  theme_minimal()
    \end{verbatim}
    \end{minipage}
    \hfill
    \begin{minipage}{0.5\textwidth}
        \centering
        \includegraphics[width=\textwidth]{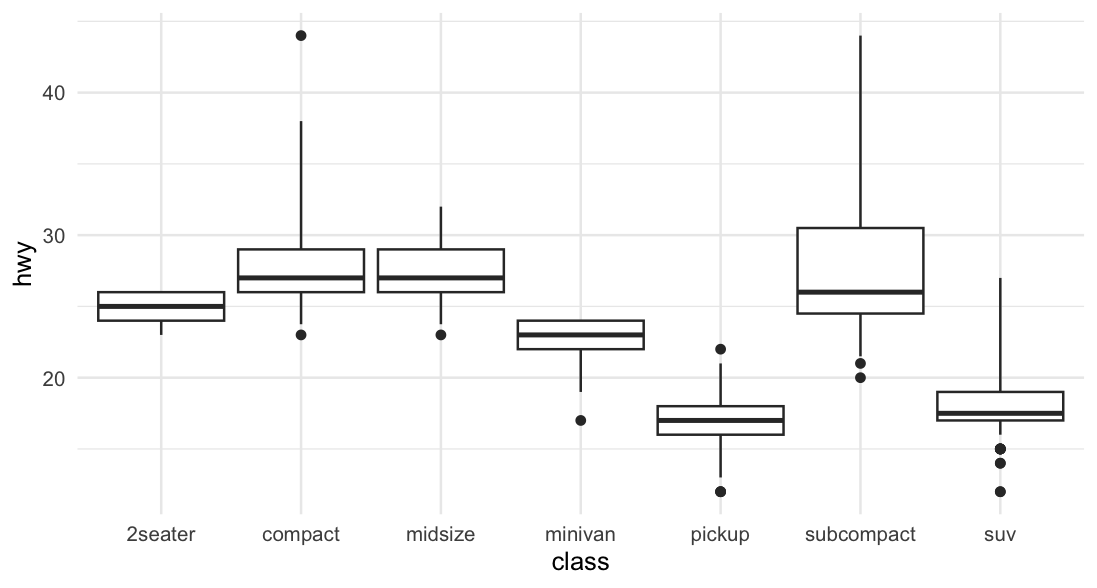}
    \end{minipage}
\end{figure}

This example replaces the default Tukey boxplot logic with the method proposed by \cite{walker2018improved}, which adjusts whisker length based on Bowley’s coefficient. It allows for improved outlier detection when data are skewed.

\subsection{Grouped Comparisons}

The function works seamlessly with grouped data, making it suitable for comparing subgroups:


\begin{figure}[ht]
    \centering
    \begin{minipage}{0.55\textwidth}
    \begin{verbatim}
> ggplot(mpg, aes(x = class, 
+                 y = hwy, 
+                 fill = as.factor(trans))) +
+  geom_skewboxplot(method = "adil") +
+  labs(fill = "") +
+  theme_minimal() + 
+  theme(legend.position = "top")
    \end{verbatim}
    \end{minipage}
    \hfill
    \begin{minipage}{0.40\textwidth}
        \centering
        \includegraphics[width=\textwidth]{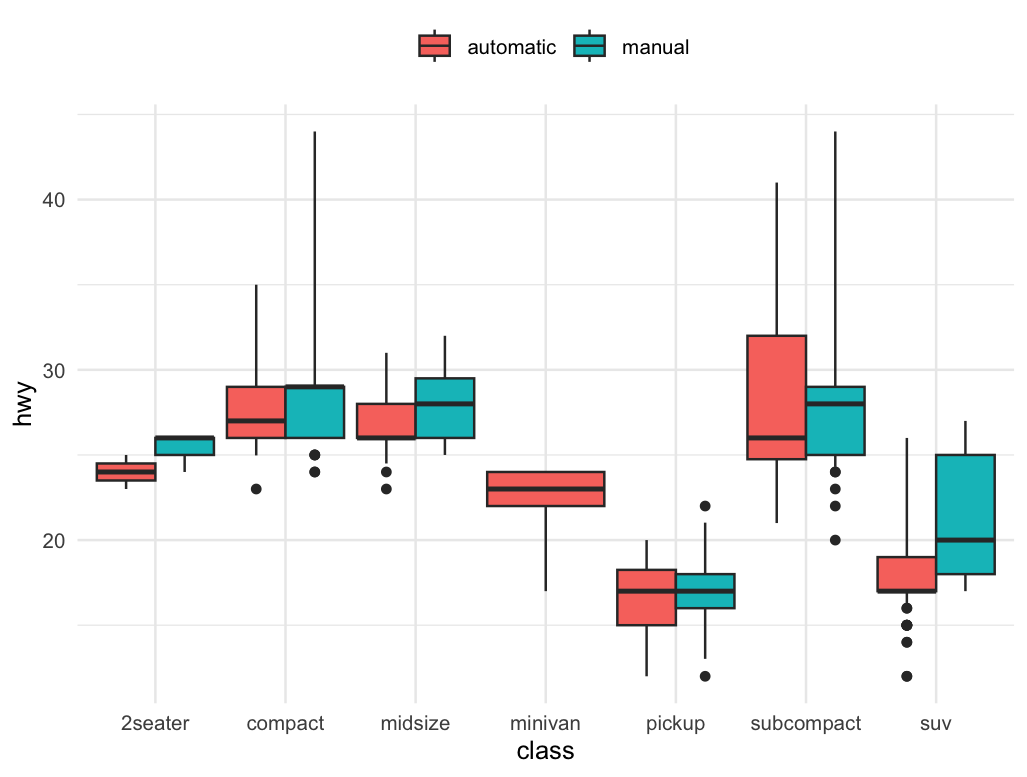}
    \end{minipage}
\end{figure}

Here, \cite{adil2015modified} method adjusts whisker extent using both skewness and the median couple, providing a more robust visualization in the presence of long-tailed distributions.

\subsection{Integration with \texttt{dplyr} and \texttt{summarise()}}

The package also includes a summarization function \texttt{summarise\_skewbox()}, useful for programmatically extracting adjusted boxplot statistics:

\begin{verbatim}
> library(dplyr)
> mpg |>
+   group_by(class) |>
+   summarise_skewbox(hwy, method = "hubert")

# A tibble: 7 × 7
  class       ymin lower middle upper  ymax outliers  
  <chr>      <dbl> <dbl>  <dbl> <dbl> <dbl> <list>    
1 2seater     24.0  24     25    26      26 <int [1]> 
2 compact     26.0  26     27    29      44 <int [10]>
3 midsize     25.9  26     27    29      32 <int [6]> 
4 minivan     21.9  22     23    24      24 <int [2]> 
5 pickup      16.0  16     17    18      22 <int [7]> 
6 subcompact  24.5  24.5   26    30.5    44 <int [9]> 
7 suv         16.9  17     17.5  19      27 <int [12]>
\end{verbatim}

This produces a tidy tibble with modified lower and upper whiskers, central tendency measures, and outlier identification, which can be used for custom visualizations or reporting.

\subsection{Parameter Sensitivity}

The tuning parameter \texttt{k} can be adjusted to control the whisker extent, allowing further customization depending on the context:


\begin{figure}[ht]
    \centering
    \begin{minipage}{0.45\textwidth}
    \begin{verbatim}
> ggplot(mpg, aes(x = class, y = hwy)) +
+   geom_skewboxplot(method = "babura", 
                     k = 2) +
+   theme_bw()
    \end{verbatim}
    \end{minipage}
    \hfill
    \begin{minipage}{0.5\textwidth}
        \centering
        \includegraphics[width=\textwidth]{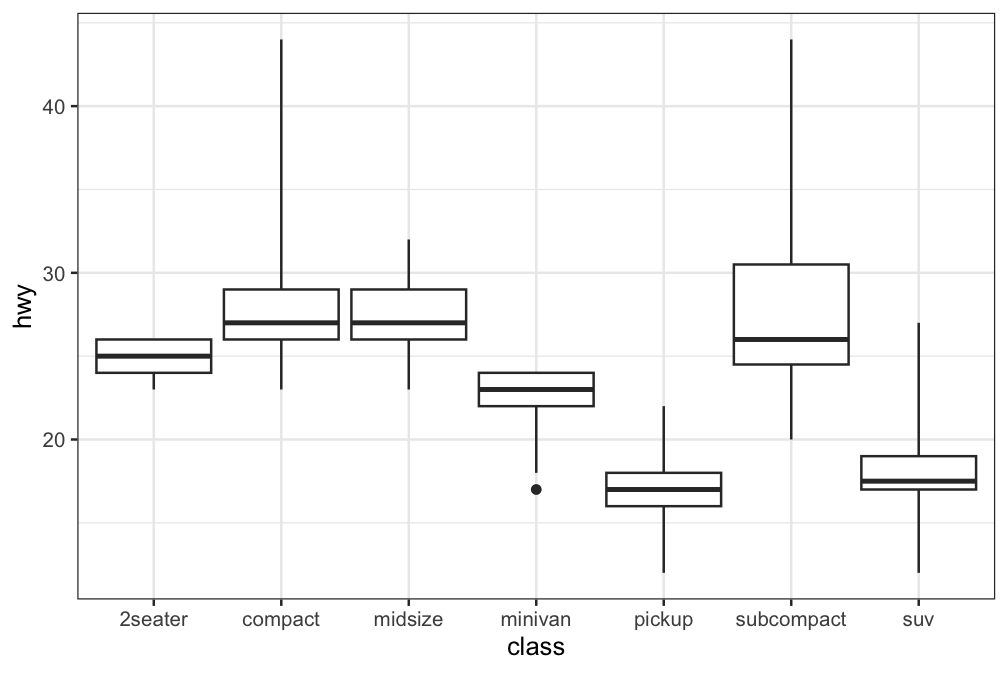}
    \end{minipage}
\end{figure}

A larger \texttt{k} reduces the number of detected outliers, while a smaller value increases sensitivity. This makes the method adaptable to domain-specific tolerance for extreme values.



\section{Simulations} 

This section presents the design and implementation of simulation studies conducted to evaluate the performance of various alternative boxplot methods under different distributional scenarios. Since traditional boxplots struggle with skewed and heavy-tailed data, we aim to assess how well the adjusted boxplots can control false detection rates of outliers across a wide range of skewness and kurtosis levels.

Simulations are based on synthetic datasets generated from the Skewed Exponential Power Distribution, which offers flexible control over skewness and tail behavior. The Mosaic Method is employed to visualize the performance across a grid of parameter combinations, providing an intuitive and comprehensive assessment framework.

\subsection{Design}

To evaluate the performance of various boxplot methods in detecting outliers under different distributional conditions, we designed two complementary simulation scenarios targeting distinct types of classification errors: \textit{swamping} and \textit{masking}. These simulations aim to assess the \textit{specificity} (\textit{true negative rate}) and \textit{sensitivity} (\textit{true positive rate}) of each method, respectively.

In the swamping scenario, datasets were generated solely from the Skewed Exponential Power Distribution (SEPD), with no artificially added outliers. This setup evaluates each method's ability to accurately identify regular observations as outliers. A high rate of false positives in this scenario indicates a tendency toward swamping — where normal data points, especially those in the skewed or heavy-tailed regions, are incorrectly flagged as outliers.

In the masking scenario, we introduced a small proportion (5\%) of synthetic outliers into the SEPD samples. These outliers were added to the extreme tails of the distribution, representing genuine anomalies. This scenario measures sensitivity by quantifying the proportion of actual outliers that each method fails to detect. A low detection rate suggests a masking effect, where true outliers are overlooked due to insufficiently adaptive whisker lengths.

We performed simulations using the Mosaic Method, detailed in the next section, on $49 \times 49$ parameter grids covering a broad range of skewness and kurtosis levels. Simulations were run for sample sizes of $n = 20, 50,$ and $100$. For each grid cell, i.e., combination of skewness and tail weight parameters, $10,\!000$ Monte Carlo replications were conducted.

Each cell in the grid represents the average proportion of misclassified observations for the given parameter combination. The color scale ranges from darker to lighter tones, corresponding to lower (0\%) to higher (10\%) misclassification rates. Darker regions thus indicate better performance, where either swamping (false positives) or masking (false negatives) is minimal. This visual representation provides an intuitive understanding of how each method responds to changes in data asymmetry and kurtosis.

\subsection{Mosaic Method}
\label{sec:mos}

The mosaic method \cite{sanchez2018visualizing} is a visual framework for summarizing the performance of statistical procedures across a wide range of distributional conditions. It works by partitioning a plot into $k\times k$ subgraphs, where each subgraph corresponds to a unique combination of parameters governing a distribution family. In our study, each subgraph represents a simulation result under a specific combination of skewness and kurtosis levels based on the Skewed Exponential Power Distribution (SEPD) \cite{zhu2009properties}. The probability density function of SEPD is defined as:

\begin{equation}
\label{eq:sepd}
f(x \mid \mu, \sigma, \alpha, p) =
\begin{cases} 
\frac{1}{K(p)} \exp\left(-\frac{1}{p} \frac{|x - \mu|^p}{2 \alpha \sigma^p}\right), & x \leq \mu \\
\frac{1}{K(p)} \exp\left(-\frac{1}{p} \frac{|x - \mu|^p}{2 (1 - \alpha) \sigma^p}\right), & x > \mu
\end{cases}    
\end{equation}

\noindent where $K(p) = \frac{1}{2} p^{1/p} \Gamma\left(1 + \frac{1}{p}\right)$ is a normalizing constant, $\mu$ is the location parameter, $\sigma$ the scale, $\alpha \in [0, 1]$ controls skewness, and $p > 0$ controls tail weight. When $\alpha = 0.5$, the distribution is symmetric; as $p \to \infty$, the distribution becomes more uniform; smaller $p$ values result in heavier tails.

\begin{figure}[h]
\centering
\includegraphics[width=0.95\textwidth]{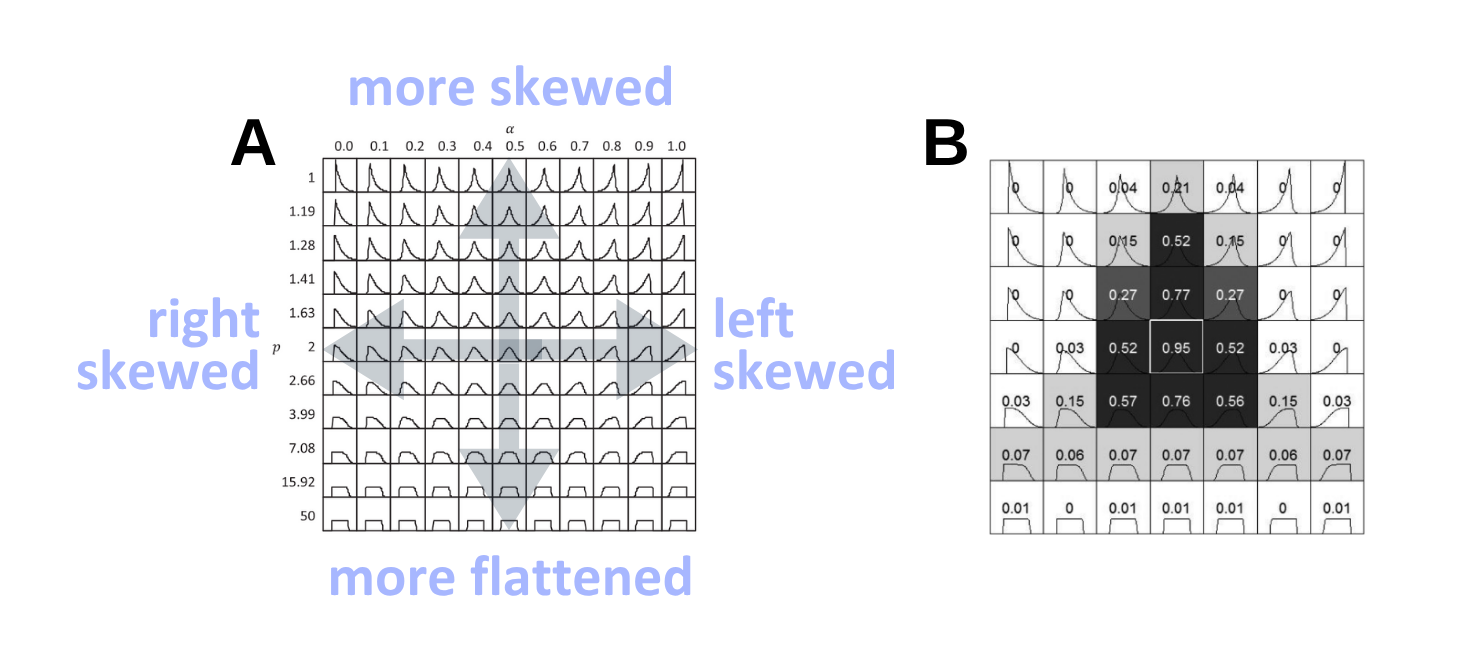}
\caption{Illustration of the mosaic approach. \textbf{Panel A}: Shape changes in SEPD as $\alpha$ and $p$ vary. It visualizes how the shape of the SEPD varies with $\alpha$ (horizontal axis) and $p$ (vertical axis). Moving horizontally corresponds to increasing skewness (from right-skewed to left-skewed), while moving vertically corresponds to increasing flatness or decreasing kurtosis. \textbf{Panel B}: Simulated performance of an outlier detection method across the same grid. It demonstrates the application of the mosaic method in our simulation study. Each tile shows the proportion of false positives or true positives (depending on the scenario) for a specific $(\alpha, p)$ pair. Darker tiles indicate better performance (lower error rates), while lighter ones indicate poorer performance.}
\label{fig:mosaic_example}
\end{figure}

This approach is illustrated in Figure~\ref{fig:mosaic_example}, which allows us to evaluate and compare the robustness of boxplot variants across a broad and continuous range of distributional shapes, rather than relying on a limited set of canonical distributions, e.g., normal, exponential, etc. It also provides a cost-effective alternative to enumerating dozens of separate simulations for each distributional family, as the SEPD can flexibly approximate many common unimodal, skewed, and heavy or light-tailed distributions.

\section{Results} 
This section presents a comparative evaluation of several boxplots under two challenging scenarios: swamping and masking. Swamping refers to the incorrect flagging of regular observations as outliers, whereas masking denotes the failure to identify true outliers hidden within the bulk of the data. For both scenarios, extensive simulations were conducted across a 49 × 49 grid of distributional settings, and the outcomes are visualized using mosaic plots. These results provide a systematic assessment of how each method responds to varying degrees of skewness, tail weight, and sample size.
\subsection{Swamping}

\begin{figure}[H]
\centering
\begin{adjustbox}{center}
\begin{tabular}{
  >{\raggedleft\arraybackslash}m{5.2cm}
  @{\hspace{0.05cm}} c
  @{\hspace{0.05cm}} c
  @{\hspace{0.05cm}} c
}
  & $n = 20$ & $n = 50$ & $n = 100$ \\

  Tukey's boxplot & 
  \includegraphics[width=0.16\textwidth]{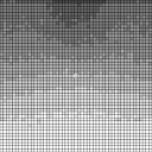} &
  \includegraphics[width=0.16\textwidth]{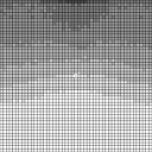} &
  \includegraphics[width=0.16\textwidth]{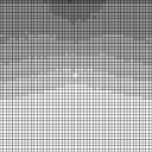} \\[-0.075cm]

  Kimber's boxplot &
  \includegraphics[width=0.16\textwidth]{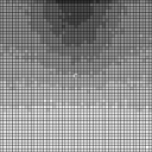} &
  \includegraphics[width=0.16\textwidth]{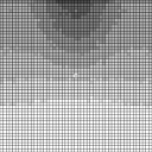} &
  \includegraphics[width=0.16\textwidth]{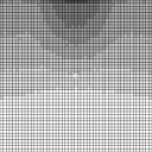} \\[-0.075cm]

  Hubert's boxplot &
  \includegraphics[width=0.16\textwidth]{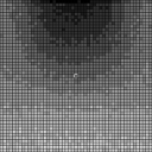} &
  \includegraphics[width=0.16\textwidth]{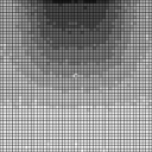} &
  \includegraphics[width=0.16\textwidth]{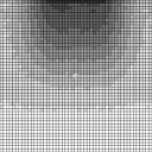} \\[-0.075cm]

  Adil's boxplot &
  \includegraphics[width=0.16\textwidth]{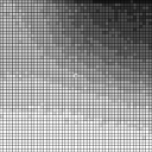} &
  \includegraphics[width=0.16\textwidth]{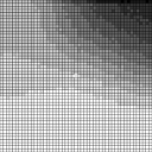} &
  \includegraphics[width=0.16\textwidth]{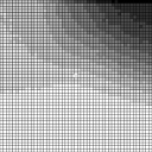} \\[-0.075cm]

  Babura's boxplot &
  \includegraphics[width=0.16\textwidth]{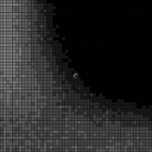} &
  \includegraphics[width=0.16\textwidth]{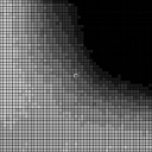} &
  \includegraphics[width=0.16\textwidth]{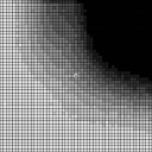} \\[-0.075cm]

  Walker's boxplot &
  \includegraphics[width=0.16\textwidth]{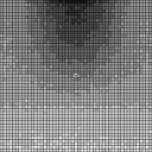} &
  \includegraphics[width=0.16\textwidth]{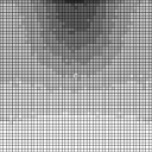} &
  \includegraphics[width=0.16\textwidth]{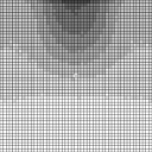} \\[-0.075cm]

  Junsawang's boxplot &
  \includegraphics[width=0.16\textwidth]{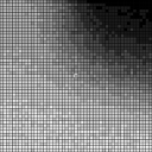} &
  \includegraphics[width=0.16\textwidth]{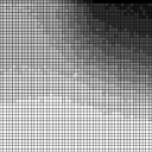} &
  \includegraphics[width=0.16\textwidth]{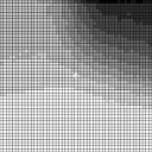} \\

\end{tabular}
\end{adjustbox}
\caption{Simulation results for \textit{swamping scenario} under the 49 x 49 mosaic plots. The colors are coded from $0\%$ to $10\%$ outliers, from darker to lighter. The grids become lighter as the proportion of falsely flagged outliers increases. In other words, it can be inferred that outlier detection performance is high where the grid colors become darker.}
\label{fig:swampling}
\end{figure}

Figure~\ref{fig:swampling} compares the performance of seven alternative boxplots against swamping for sample sizes of $n=20$, $n=50$, and $n=100$. Tukey's boxplot shows that it tends to incorrectly label non-outlying data as outliers under almost all distributional conditions, especially in skewed and platykurtic distributions. Its performance does not show a notable improvement as the sample size increases. 

Kimber's boxplot shows a slight improvement over Tukey but still has light colors, particularly in the regions of extreme skewness (left and right edges of the mosaics) and flat distributions (bottom parts of the mosaics). This suggests that despite addressing asymmetry with semi-interquartile ranges, the method does not fully resolve the swamping problem in skewed distributions. 

Hubert's boxplot features darker tones towards the center of the mosaics. This indicates that thanks to the medcouple statistic, it successfully lowers the false labeling rate in symmetric or moderately skewed distributions. However, the colors become lighter in the corners and bottom regions corresponding to extremely skewed and flat distributions, indicating an increased risk of swamping under these extreme conditions. 

A modification of Hubert's boxplot, Adil's boxplot displays a pattern similar to Hubert's across the mosaics. It performs well in symmetric and moderately skewed distributions. However, the presence of a darker area in the top-left corner for $n = 20$ suggests that it might be more successful under conditions of leptokurtic and right-skewed distributions, even with small sample sizes. 

Babura's boxplot exhibits very dark tones across a large portion of the mosaic plot for n=20, particularly even in the extremely skewed bottom-left and bottom-right corners (flat distributions). This indicates that using Bowley's coefficient of skewness, it is highly resistant to swamping, even with small sample sizes, making it one of the best-performing methods. Although these dark areas narrow slightly as the sample size increases ($n = 50$ and $n = 100$), their overall performance remains very strong. 

Like Babura's boxplot, Walker's boxplot also uses Bowley's coefficient. The plots for this method have dark tones in the bottom-left and bottom-right regions of the mosaics. This reveals that it performs very well in extremely skewed and platykurtic distributions. It generally displays a strong and consistent performance, similar to Babura's boxplot. 

Junsawang's boxplot, at the bottom, also shows good performance with dark tones in the lower parts of the mosaics (platykurtic distributions) and in the skewed regions. By dynamically adjusting Hubert's approach with the ratio of semi-interquartile ranges, the method proves to be resistant to the swamping problem across a broad spectrum of distributions.

We can conclude that the performance of Tukey's boxplot is inadequate for skewed and heavy-tailed distributions, as they tend to produce a high rate of false positives, a phenomenon known as swamping. The simulation results, visualized through mosaic plots, consistently show that Tukey's boxplot is highly susceptible to misclassifying normal observations as outliers across a wide range of distributional shapes. In stark contrast, several alternative methods, including Babura's, Walker's, and Junsawang's boxplots, demonstrate a significantly improved ability to handle these challenging distributional conditions. These methods, which incorporate robust measures of skewness like Bowley's coefficient or adjusted interquartile ranges, successfully reduce the proportion of falsely identified outliers, resulting in a more accurate and reliable delineation of extreme values. 

\subsection{Masking}

\begin{figure}[H]
\centering
\begin{adjustbox}{center}
\begin{tabular}{
  >{\raggedleft\arraybackslash}m{5.2cm}
  @{\hspace{0.05cm}} c
  @{\hspace{0.05cm}} c
  @{\hspace{0.05cm}} c
}
  & $n = 20$ & $n = 50$ & $n = 100$ \\

  Tukey's boxplot & 
  \includegraphics[width=0.16\textwidth]{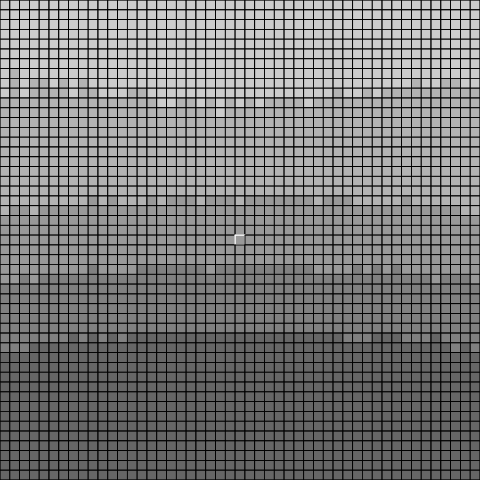} &
  \includegraphics[width=0.16\textwidth]{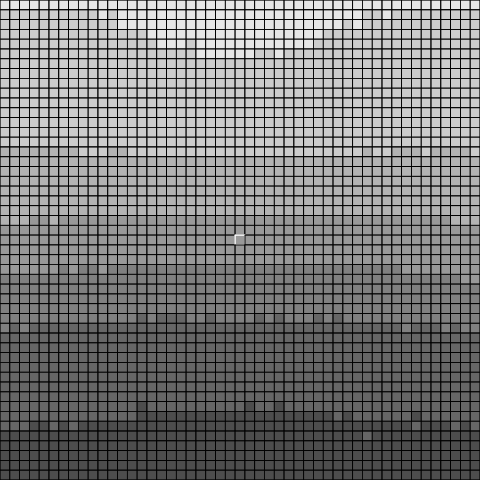} &
  \includegraphics[width=0.16\textwidth]{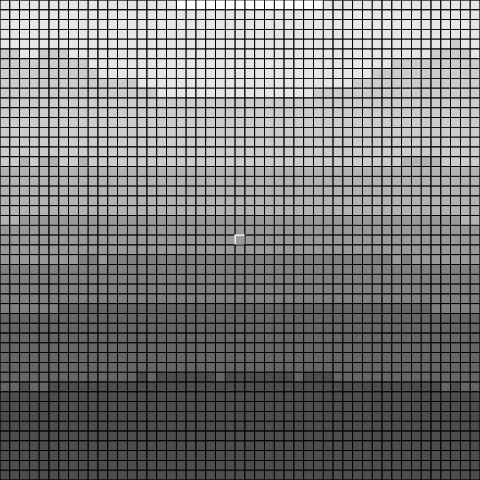} \\[-0.075cm]

  Kimber's boxplot &
  \includegraphics[width=0.16\textwidth]{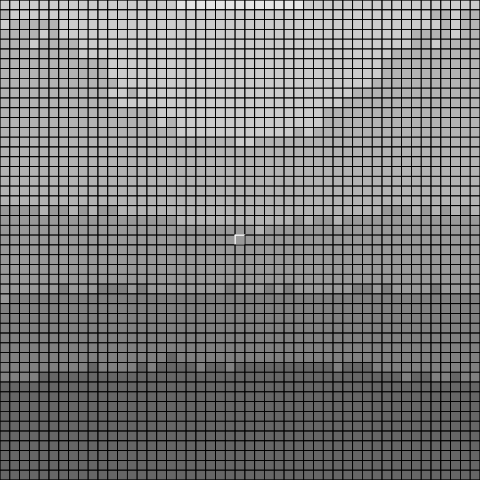} &
  \includegraphics[width=0.16\textwidth]{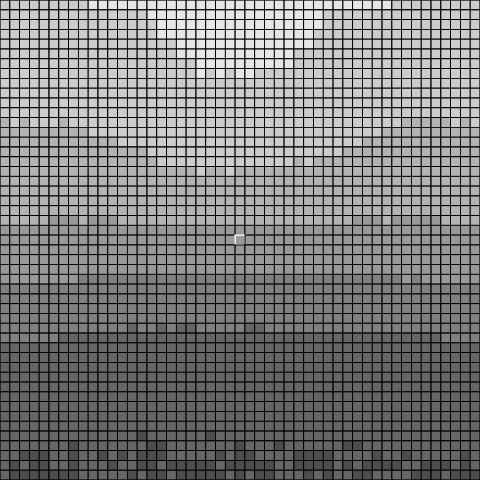} &
  \includegraphics[width=0.16\textwidth]{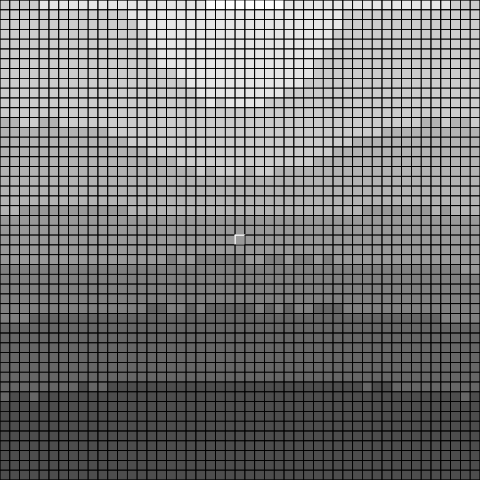} \\[-0.075cm]

  Hubert's boxplot &
  \includegraphics[width=0.16\textwidth]{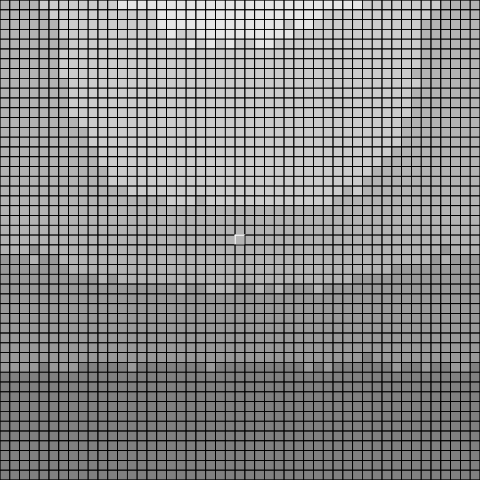} &
  \includegraphics[width=0.16\textwidth]{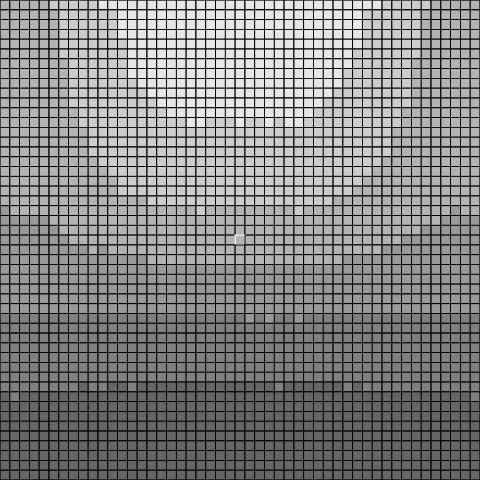} &
  \includegraphics[width=0.16\textwidth]{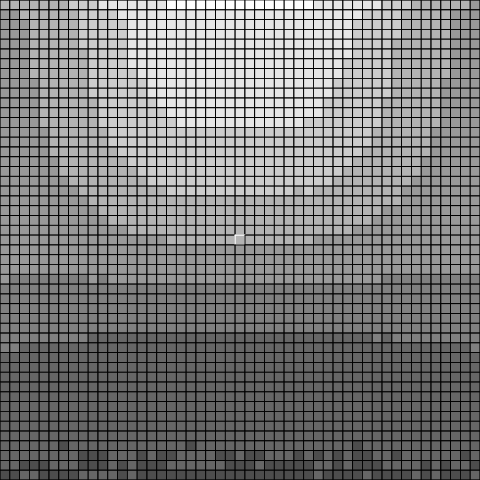} \\[-0.075cm]

  Adil's boxplot &
  \includegraphics[width=0.16\textwidth]{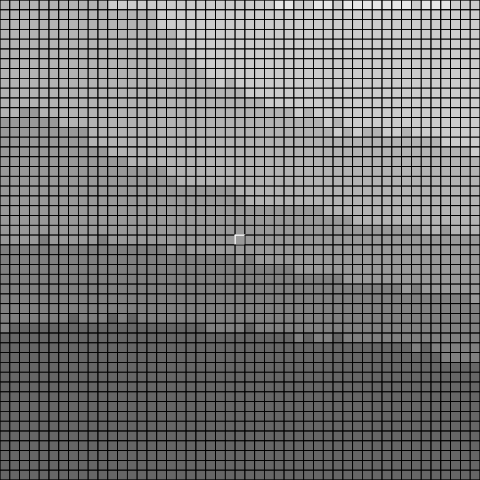} &
  \includegraphics[width=0.16\textwidth]{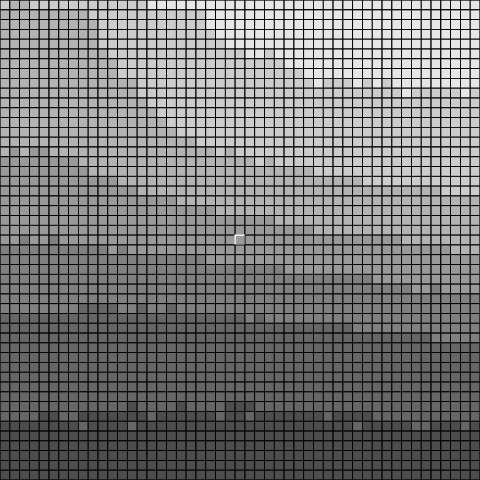} &
  \includegraphics[width=0.16\textwidth]{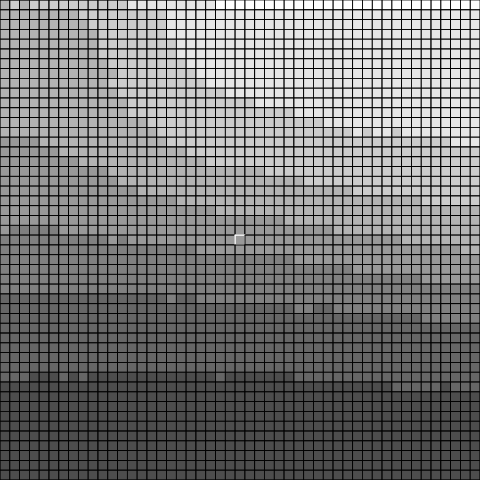} \\[-0.075cm]

  Babura's boxplot &
  \includegraphics[width=0.16\textwidth]{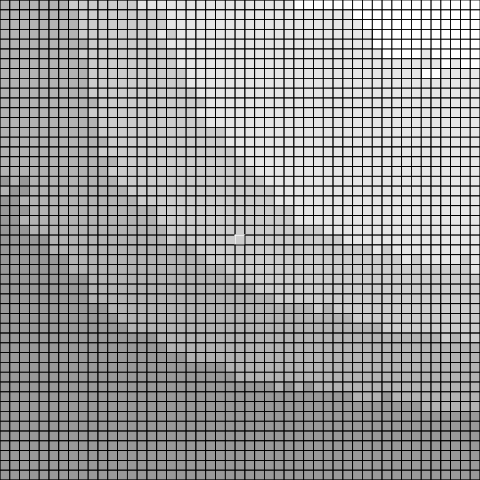} &
  \includegraphics[width=0.16\textwidth]{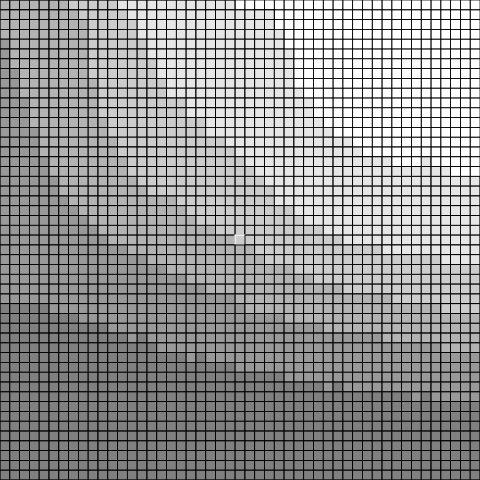} &
  \includegraphics[width=0.16\textwidth]{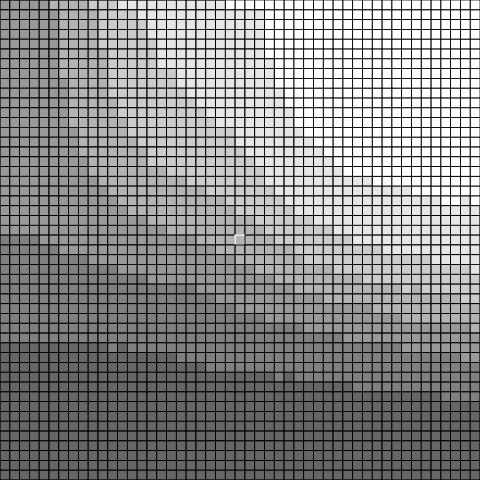} \\[-0.075cm]

  Walker's boxplot &
  \includegraphics[width=0.16\textwidth]{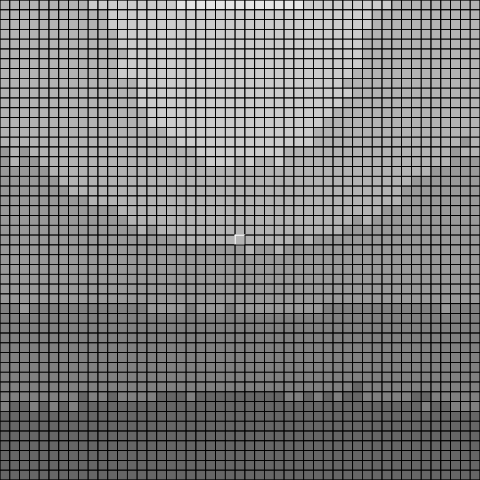} &
  \includegraphics[width=0.16\textwidth]{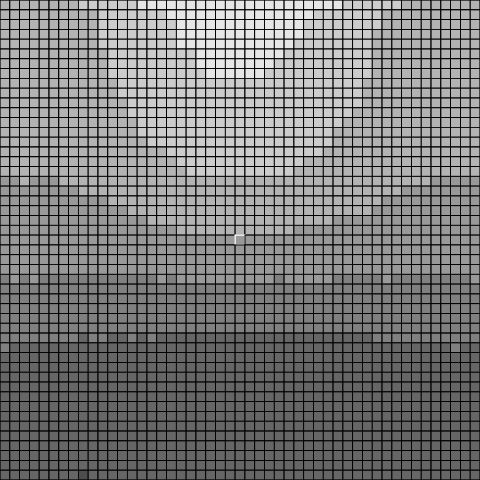} &
  \includegraphics[width=0.16\textwidth]{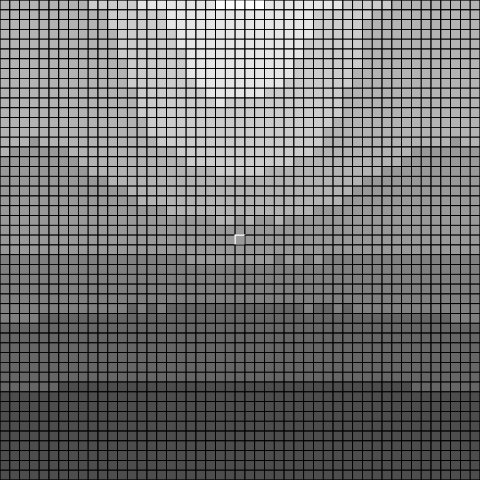} \\[-0.075cm]

  Junsawang's boxplot &
  \includegraphics[width=0.16\textwidth]{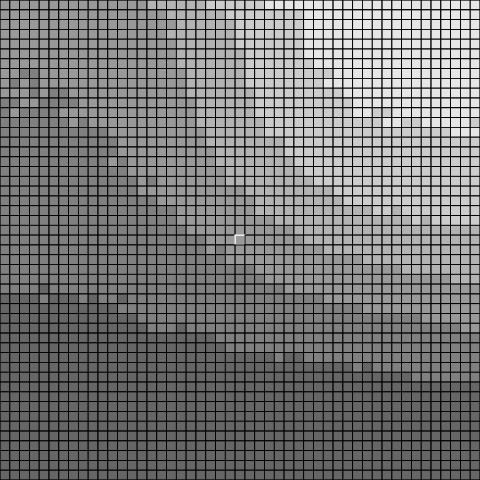} &
  \includegraphics[width=0.16\textwidth]{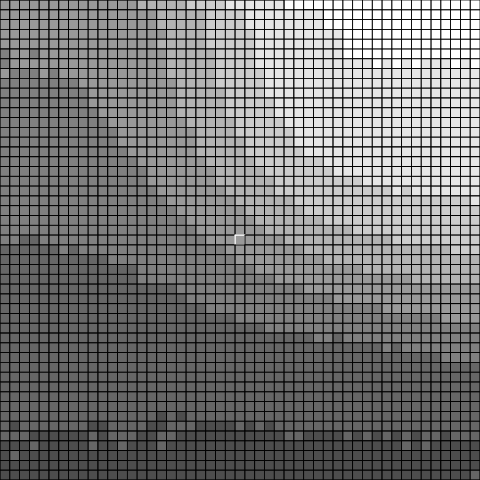} &
  \includegraphics[width=0.16\textwidth]{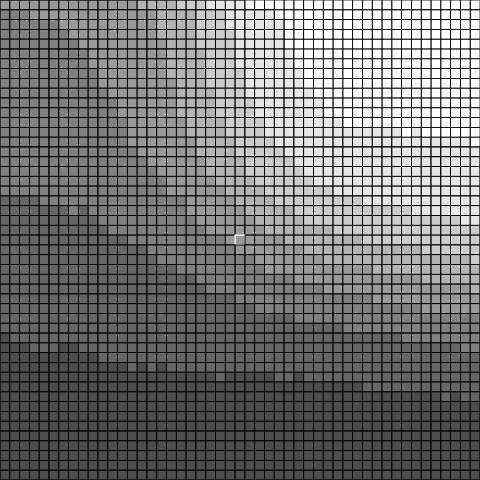} \\

\end{tabular}
\end{adjustbox}
\caption{Simulation results for the \textit{masking} scenario under the 49 × 49 mosaic plots. The colors are coded from 0\% to 10\% masking severity, transitioning from darker to lighter tones. Each cell reflects the proportion of undetected contaminated observations masked by outliers. Lighter grid cells indicate a higher level of masking, meaning that a larger portion of outliers remains hidden within the bulk of the data, while darker cells represent parameter combinations where outlier detection methods perform more effectively, revealing fewer masked outliers.}
\label{fig:masking}
\end{figure}

Figure~\ref{fig:masking} compares the performance of seven alternative boxplots against masking for $n = 20$, $n = 50$, and $n = 100$, where the focus is on the proportion of actual outliers that remain undetected. Darker cells in the mosaic plots correspond to lower masking severity, i.e., better detection of true outliers, whereas lighter cells indicate a higher proportion of undetected outliers hidden within the bulk of the data.

Tukey's boxplot exhibits predominantly light colors across all sample sizes and parameter combinations. This pattern indicates that the traditional boxplot is highly susceptible to masking, frequently failing to detect true outliers when the data are skewed or heavy-tailed. Increasing the sample size does not substantially improve its sensitivity; even at $n = 100$, the method shows a considerable inability to detect anomalous observations, especially in regions corresponding to extreme skewness or low tail thickness. Kimber's boxplot performs slightly better than Tukey, as reflected by somewhat darker areas in the central and moderately skewed regions of the mosaics. Nevertheless, it still displays light tones in the highly skewed or platykurtic areas, indicating difficulties in distinguishing true outliers from the main data mass. Although Kimber's use of semi-interquartile ranges aids in handling asymmetry, it remains insufficient for capturing extreme tail behavior.

Hubert's boxplot demonstrates a clear improvement over both Tukey and Kimber in reducing masking. The mosaics show darker tones around the central and moderately skewed regions, highlighting the value of the medcouple adjustment in enhancing sensitivity. However, as skewness and flatness increase toward the edges and bottom of the mosaic, the colors lighten, indicating that the method still struggles to detect outliers when the distribution deviates strongly from symmetry. Adil's boxplot displays a pattern similar to Hubert's boxplot, with strong performance in symmetric or moderately skewed scenarios. In certain regions---particularly at $n = 20$---the presence of darker shades in the upper-left area suggests that the combined use of medcouple and classical skewness enhances the detection of outliers in leptokurtic and right-skewed distributions. Nonetheless, sensitivity declines in extremely skewed or highly platykurtic scenarios.

Babura's boxplot stands out with consistently dark regions throughout large portions of the mosaics, even at $n = 20$. This indicates a robust capability to detect true outliers across a wide spectrum of distributional shapes. Only minor lightening appears in the extreme corners, suggesting a slight drop in performance under conditions of both extreme skewness and very flat tails. However, Babura’s overall resistance to masking is particularly for small sample sizes.

Walker's boxplot shows a performance profile similar to Babura's boxplot, which is expected given their shared reliance on Bowley’s skewness. The method maintains dark tones in most regions of the mosaic across all sample sizes, especially in highly skewed and platykurtic conditions. This suggests that using the ratio-based adjustment for Bowley’s coefficient strengthens the method’s ability to expose true outliers.

Junsawang's boxplot also demonstrates resistance to masking, with dark tones concentrated in the skewed and flat regions. This highlights the value of dynamically adapting Hubert's boxplot using the ratio of semi-interquartile ranges. By reflecting local asymmetry around the median, this method successfully detects outliers in many scenarios where other approaches fail.

Overall, the results reveal that Tukey's boxplot is inadequate when subtle outliers are embedded in skewed or heavy-tailed data. In contrast, methods incorporating robust skewness measures---particularly those based on Bowley’s coefficient, such as Babura's and Walker's, as well as the dynamic adjustment of Junsawang's boxplots---demonstrate superior sensitivity. These methods maintain low masking severity across a broad range of distributional shapes and sample sizes, making them strong candidates for applications where capturing true extreme observations is essential.

\section{Conclusions} 
This paper introduced \texttt{ggskewboxplots}, which integrates a comprehensive collection of skewness-aware boxplot methods for visualizing asymmetric and heavy-tailed distributions. By embedding multiple robust whisker-adjustment strategies into the familiar grammar of graphics, the package provides a practical and unified framework for improving outlier detection in exploratory data analysis.

Extensive Monte Carlo simulations conducted under controlled skewness–kurtosis conditions revealed clear performance differences among the examined boxplot variants. The traditional Tukey-style boxplot consistently exhibited high rates of swamping and masking, especially in highly skewed or platykurtic settings, demonstrating its limited reliability for non-symmetric data. Methods incorporating quartile-based skewness measures, such as Babura's and Walker's boxplots, showed the strongest and most consistent performance across both swamping and masking scenarios, retaining robustness even at small sample sizes. Similarly, the Junsawang's boxplot performed well in complex distributional structures by adapting whisker lengths to local asymmetry around the median. Adjusted approaches relying on the medcouple—including those of Hubert's and Adil's boxplots—also improved performance relative to classical methods, particularly in moderately skewed distributions.

Overall, the simulation results underscore that no single boxplot method is uniformly optimal, but robust skewness-aware variants generally provide substantial gains in sensitivity and specificity under asymmetric distributions. The \texttt{ggskewboxplots} package translates these methodological improvements into an accessible visualization toolkit, allowing practitioners to select appropriate boxplot forms for their data without leaving the \texttt{ggplot2} workflow. Future work may extend the package to multivariate analogues or integrate automatic diagnostic tools that recommend suitable boxplot methods based on empirical skewness and kurtosis. Together, these contributions enhance the reliability of visual exploratory analysis in real-world settings where deviation from symmetry is the norm rather than the exception.

\subsection*{Acknowledgment}

The preliminary work \cite{cavus2025} of this paper was presented at the \hyperlink{https://dagstat2025.de/}{7th Joint Statistical Meeting of the Deutsche Arbeitsgemeinschaft Statistik 2025}. This paper is financially supported by the Eskisehir Technical University Scientific Research Projects Commission under grant no. 23ADP050.


\end{document}